\documentclass[aps,onecolumn,10pt,nofootinbib]{revtex4}
\usepackage{graphicx,amssymb,amsmath,subfigure}

\newcommand{\para}[1]{\left(#1\right)}
\newcommand{\abs}[1]{\left|#1\right|}

\begin{document}
\title{Gauge theory of the Hubbard model on honeycomb lattice
and its instanton effect}
\author{Abolhassan Vaezi$^{*}$ and Xiao-Gang Wen}
\affiliation{Department of Physics,
Massachusetts Institute of Technology,
Cambridge, MA 02139, USA}
\email{vaezi@mit.edu}

\date{\today}

\begin{abstract}
In this paper we investigate possible spin disordered phase in the Hubbard
model on the honeycomb lattice.  Using a slave-particle theory that include the
charge fluctuations, we find a meanfield  spin disordered phase in a range of
on-site repulsion $U$.  The spin disordered state is described by gapped
fermions coupled to compact U(1) gauge field.  We study the
confinement/deconfinement problem of the U(1) gauge theory due to the
instantons proliferation. We calculate all allowed instanton terms and compute
their quantum numbers. It is shown that the meanfield spin disordered phase is
unstable. The instantons proliferation induce a translation symmetry breaking.
\end{abstract}
\maketitle

\section{Introduction}

Hubbard model\cite{Hubbard_1963,Hirsch_1985a} is believed to describe the
physics of many strongly correlated systems {\em e.g.}, Mott
insulator\cite{Mott_1949,Imada_et_al_1998} and high temperature
superconductors\cite{Bednorz_Mueller_1986,Anderson_1987Sci,Lee_Nagaosa_Wen_2006a}.
Motivated by the experiments of high temperature
superconductors, people have been look for
spin liquid phase\cite{RS9173,W9164,Wen_2002c} that does not break any
symmetry in in large $U$ Hubbard model and its generalization.

Recently, Meng {\em et al.} \cite{Meng_2010a,Furukawa_1992,Lilly_1990} have
studied the Hubbard model on the honeycomb lattice using quantum Monte Carlo
(QMC) method. They only consider the bipartite system in which the Hamiltonian
only connects states on different sublattices. The reason is that the QMC does
not have sign problem in this case.  They report a gapped spin liquid phase for
a range of U/t. This phase is an insulating that does not appear to break any
symmetry. The
true nature of the phase is still under debate.

Recently, in Ref. \cite{Vaezi_2010a}, we have studied the phase diagram of the
Hubbard model on the honeycomb lattice (see Fig. 1 and 2) using the generalized
slave-particle technique
\cite{Anderson_Zou_a,Florens_Georges_2004a,PA_Lee_SS_Lee_2004a,Senthil_2008_a}
which include charge fluctuations.  Within meanfield approximation, we find a
Mott phase transition to the insulating phase at $U_{c1}\simeq 2.2t$ above
which charge gap opens up and we obtain the gapped spin liquid phase
(spin/charge gapped phase). There is also another phase transition between the
spin liquid and the anti-ferromagnetic order phases at $U_{c2}\simeq 3t$. In
this phase, in contrast to the the gapped spin liquid phase, the mass of
spinons is very small and negligible. In the case of nearest neighbor hopping
Hubbard model, which is a bipartite system, the gauge theory of our meanfield
spin liquid phase is the compact staggered U(1). In this phase, all excitations
are gapped except gauge fluctuations.  These results are within mean-field and
due to the compactness of the U(1) lattice gauge theory, stability of such
mean-field states are under question.  In compact $U(1)$ gauge theories,
instanton (anti-instanton) configurations are allowed and when they
proliferate, spinons become confined and the results of the mean-field are no
longer valid. Therefore studying the fate of this gapped spin liquid is
necessary.

In this paper we find that instanton configurations are relevant.
instantons have nonzero fugacity and we do obtain a confined phase.  More
importantly, instanton operators carry a non-trivial crystal momentum.  Also,
under 60 degree lattice rotation and parity, an instanton is changed to an
anti-instanton.  However, the instantons carry trivial quantum numbers for
other symmetries.  Since a triple instanton carries trivial quantum numbers for
all symmetries, so triple instanton can proliferate which leads to a confined
phase.  Since single instanton carries non-trivial crystal momentum, this
allows us to conclude that the $U(1)$ confined phase is a phase that break
translation symmetry but not spin rotation symmetry. Therefore we finally
obtain an insulating phase at half filling that breaks the lattice translation
symmetry!  On the other hand, in the presence of second neighbor hopping in the
Hubbard model the charge/spin gapped phase can be spin liquid that do not break
translation, parity, 60 degree lattice rotation, and  spin rotation symmetries.

\begin{figure}
  \includegraphics[width=180pt]{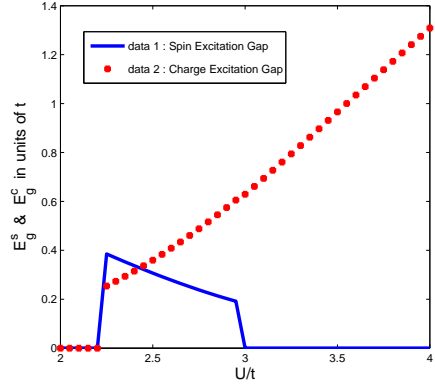}\\
  \caption{ Spin excitation gap (blue line) and charge excitation gap (red dots). Figure has been taken from Ref. \cite{Vaezi_2010a}}\label{fig1}
\end{figure}

\begin{figure}
  \includegraphics[width=180pt]{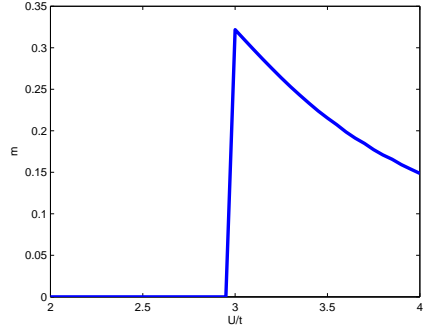}\\
  \caption{Staggered magnetization, $m$
($(-)^{i}~S_z\para{i}$), as a function of $\frac{U}{t}$.
There is a phase transition to the anti-ferromagnetic order
at $U/t=3$. After Ref. \cite{Vaezi_2010a}}\label{fig2}
\end{figure}

\section{Symmetry transformations on the honeycomb lattice}

\begin{figure}
\centering
\subfigure{
\includegraphics[height=130pt]{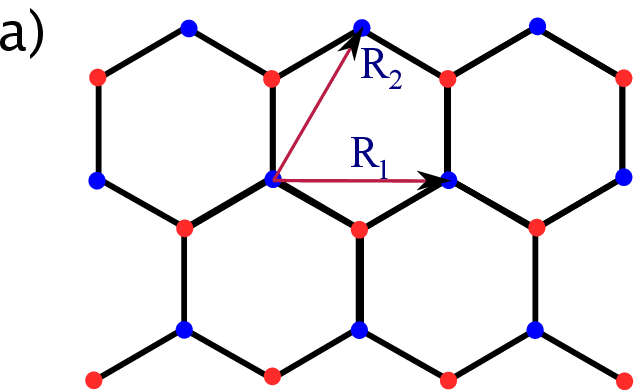}
\label{fig:subfig3a}
}
\subfigure{
\includegraphics[height=130pt]{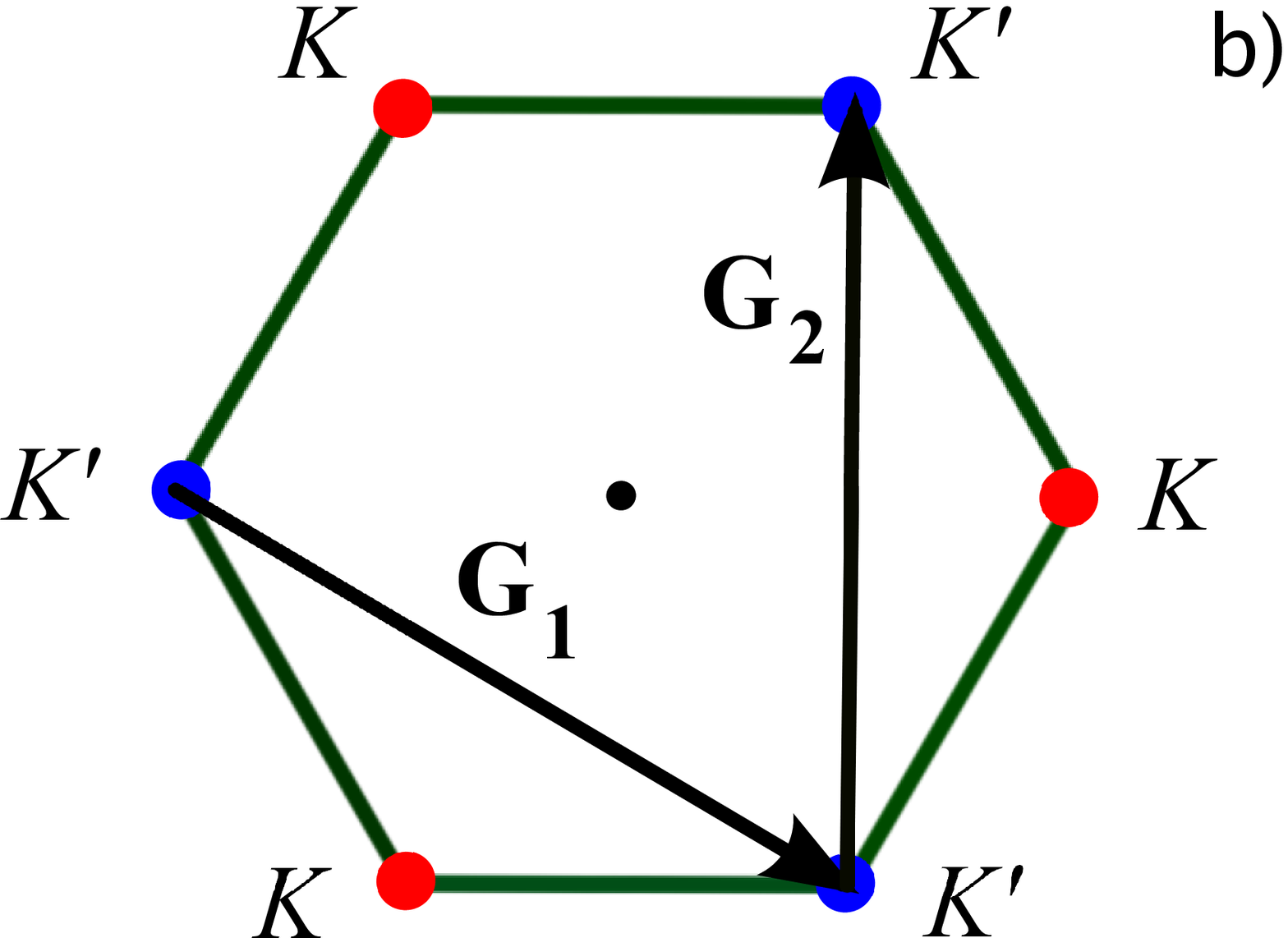}
\label{fig:subfig3b}
}
\label{fig:subfigureExample}
\caption[Optional caption for list of figures]{Honeycomb lattice and basis vectors of reciprocal lattice. \subref{fig:subfig3a}, Honeycomb lattice in real space. Blue dots represent atoms on sublattice A and the red dots represent atoms on sublattice B. $\vec{R}_1$ and $\vec{R}_2$ are basis vectors of the honeycomb lattice. \subref{fig:subfig3b}, Brillouin zone. $\vec{G}_1$ and $\vec{G}_2$ are basis vectors of reciprocal lattice. In the absence of $\lambda$, energy dispersion has two inequivalent nodal points at $\vec{K}$ and $\vec{K}'$. It is clear form this figure that under $C_6$ (60 degrees rotation) $K\to K'$ and $K'\to K$, while they do not change under parity $\sigma$.}\label{fig3}
\end{figure}

Since the unit cell of the honeycomb lattice has two sites in it, and we can
label them by A and B or $s=0,1$, any lattice point can be represented as
$\vec{R}=x_1 \vec{a}_1+x_2 \vec{a}_2+s \hat{y} \equiv \para{x_1,x_2,s}$, where
$\vec{a}_1=\sqrt{3}a\para{1,0}$ (a is the lattice spacing between A and B
atoms), $\vec{a}_2=\sqrt{3}a\para{\frac{1}{2},\frac{\sqrt{3}}{2}}$ and $s$ takes 0 and 1
values. The basis vectors of the reciprocal lattice are
$\vec{G}_{1}=\frac{4\pi}{3a}\para{\frac{\sqrt{3}}{2},1}$ and $\vec{G}_{2}=\frac{4\pi}{3a}\para{0,1}$ (see Fig. 3). 
Type A atoms are connected to type B
atoms by the following three vectors: $\vec{b}_{1}=a\para{0,1}$,
$\vec{b}_{2}=a\para{\frac{\sqrt{3}}{2},\frac{-1}{2}}$ and
$\vec{b}_{3}=a\para{\frac{-\sqrt{3}}{2},\frac{-1}{2}}$. Honeycomb lattice is
invariant under five symmetry transformations: time reversal, parity
$(\sigma: \para{x,y}\rightarrow \para{x,-y})$, 60 degree rotation ($C_6$), translation
along $\vec{a}_1$ and $\vec{a}_2$ ($T_1$ and $T_2$). It is easy to check these
symmetry operations act on the lattice as the following \cite{Ying_2010_1}:
\begin{eqnarray}
 &&T_1~:~~~\para{x_1,x_2,s} ~~~~\rightarrow~~~~\para{x_1+1,x_2,s}\\
 &&T_2~:~~~\para{x_1,x_2,s} ~~~~\rightarrow~~~~\para{x_1,x_2+1,s}\\
 &&T~~:~~~\para{x_1,x_2,s} ~~~~\rightarrow~~~~\para{x_1,x_2,s}\\
 &&\sigma~~:~~~\para{x_1,x_2,s} ~~~~\rightarrow~~~~\para{x_1+x_2,-x_2,1-s}\\
 &&C_6:~~~\para{x_1,x_2,s} ~~~~\rightarrow~~~~\para{1-s-x_2,x_1+x_2+s-1,1-s}.
\end{eqnarray}

\section{Method}

In the Anderson-Zou slave particle method, electron operators are represented as:
\begin{eqnarray}
  C_{i,\sigma}^\dag= f_{i,\sigma}^\dag h_{i}+\sigma ~ d_{i}^\dag f_{i,-\sigma}=[\begin{array}{cc}
                 h_i &
                 d_i^\dag
               \end{array} ] \left[ \begin{array}{c}
                             f_{i,\sigma}^\dag \\
                             \sigma f_{i,-\sigma}
                           \end{array}\right],
\end{eqnarray}
where $f_{i,\sigma}^\dag$ creates a state with a single electron on it (a
spinon), $h_{i}^\dag$ creates a state with no charge on it (a holon), and
$d_{i}^\dag$ creates a state with two electron on site $i$ (a doublon). It
should be mentioned that the physical Hilbert space contains only four states:
empty state (holon), one electron (spinon) and two electrons (doublon) on each
site. So we always have one and only one slave particle on each site. So we
conclude that we should put the local constraint
$n_{i}^{h}+n_{i,\uparrow}^{f}+n_{i,\downarrow}^{f}+n_{i}^{d}=1$, to get rid of
redundant states. This is the physical constraint which should be satisfied on
every site. We could also obtain this result by noting that the electron
operators are fermion and should satisfy the anticommutation relations. From
the definition of $ C_{i,\sigma}^\dag$ it is obvious that it is invariant under
U(1) gauge transformation (We require $h_{i}$ and $d_{i}$  to remain bosonic
operators i.e., preserve their statistics after transformation, otherwise we
would have SU(2) gauge invariance. However at $U=\infty$ we have only fermions
and only in that case we have SU(2) gauge symmetry). It is worth noting that
all the slave particles carry the same charge under the internal U(1) gauge.
Since the above constraint and as a result the Hubbard Hamiltonian are also
gauge invariant, so is the action of the Hubbard model.

Using this slave technique the Hubbard Hamiltonian can be rewritten as the following:
\begin{eqnarray}
H=&&\sum U d_{i}^\dag d_{i}-t\sum_{<i,j>}\para{\chi_{i,j}^{f}\chi_{j,i}^{b}+\Delta_{i,j}^{f \dag}\Delta_{i,j}^{b }+h.c.}~~~~\cr
&&   +\lambda(f_{i,\uparrow}^\dag f_{i,\uparrow} +f_{i,\downarrow}^\dag f_{i,\downarrow}+h_i^\dag h_i+d_i^\dag d_i -1),
\end{eqnarray}
in which we have used these notations
$\chi_{i,j}^{f}=\sum_{\sigma}f_{i,\sigma}^\dag
f_{j,\sigma}~,~\chi_{i,j}^{b}=h_{i}^\dag h_{j}-d_{i}^\dag
d_{j}~,~\Delta_{i,j}^{f}=\sum_{\sigma}\sigma f_{-\sigma,i}
f_{j,\sigma}~,~\Delta_{i,j}^{b}= d_{i}h_{j}+h_{i}d_{j}$ . Within mean field and
by using Hubbard-Stratonovic we can decouple spinons and bosons and obtain a
mean field state.  In our numerical studies we have obtained three phases. At
small U/t limit we obtain a semi-metallic phase. At large U/t limit we obtain
AF order and for moderate values of U/t we obtain a spin liquid phase.

\section{Instanton proliferation and confinement}

Now let us focus on the spin liquid phase. In this phase: $\chi_{i,j}^{f,h}=0$.
Therefore the effective Hamiltonian of spinons in this phase is of the
following forms:
\begin{eqnarray}
  H_{s}=\lambda\sum_{i,\sigma,\tau} f_{i,\tau,\sigma}^\dag f_{i,\tau,\sigma} -t\sum_{<i,j>,\sigma,\tau}\Delta_{b}\para{i,j}\sigma f_{i,A,\sigma}^\dag f_{j,B,-\sigma}^\dag +h.c.~.
\end{eqnarray}

Now let us use the following ansatz:
$\Delta_{b}\para{\vec{\delta}}=\Delta_{b}$. So we have
$\Delta_{b}\para{\vec{k}}=\Delta_{f}\eta(\vec{k})$, where $\eta(\vec{k})=
e^{-ik_y}+2e^{+i\frac{ky}{2}}cos{\frac{\sqrt{3}}{2}k_x}$ is the structure
factor of the honeycomb lattice. Therefore the energy spectrum of spinons are:
$\sqrt{\lambda^2+\left|t\Delta_{b}\para{k}\right|^2}$. From the energy spectrum
we see that spinons are gapped. But what if we include the effect of
instantons? To answer this important question, we first study the gauge theory
of this mean-field state.

In the above effective Hamiltonian we transform operators as:
$f_{i,A,\sigma}\rightarrow e^{i\alpha}f_{i,A,\sigma}$ and
$f_{i,B,\sigma}\rightarrow e^{-i\alpha}f_{i,B,\sigma}$ for any arbitrary phase
$\alpha$, {\em i.e.} assuming a staggered global gauge transformation, then the
effective Hamiltonian does not change. Therefore the $IGG$ of the Hamiltonian
is staggered $U(1)$. The reason is that there is no hopping term due to the
non-zero charge gap and the if we the gauge transformation of two neighboring
sites have opposite phases, the total phase change of the pairing term becomes
zero and therefore gauge fluctuations are described by staggered compact $U(1)$
instead of compact $U(1)$ gauge theory. This is equivalent to assuming have
positive unit charge on sublattice A and negative unit charge on sublattice B
for slave particles under the internal gauge transformation. So, at mean field
level, the charge/spin gapped phase has a neutral spinless $U(1)$ gapless mode
as its only low energy excitations.  However, it is well known that $U(1)$
theory in 2+1D is confined due to instanton effects.  So in the latter part of
this paper, we will assume that the $U(1)$ fluctuations are weak and use the
semiclassical approach to study the $U(1)$ confined phase where the $U(1)$ mode
is gapped.

We like to remark that
it is possible to break this staggered compact $U(1)$ down to a $Z_2$ \cite{Ying_2010b,Xu_2010a,Kim_2010a,Kou_2010a,Sachdev_2010a} one by
Anderson-Higgs mechanism. If we add second neighbor hopping to the Hubbard
model, within slave boson this term generate pairing terms of the form
$f_{i,\tau,\sigma}^\dag f_{j,\tau,-\sigma}^\dag$, {\em i.e.} it induces the
same sublattice pairing and the Hamiltonian is no longer invariant under the
staggered global $U(1)$ gauge transformation. In this case gauge fluctuations
are gapped and thus our mean filed state is stable and we can trust our
results. Therefore for this case spin liquid phase is physical. On the
frustrated lattices like the triangular lattice the gauge theory is $Z_2$
because we cannot dive the lattice in two sublattices, A and B.

It is useful to do particle hole transformation on sublattice $B$, so that we
can see the gauge theory of the transformed Hamiltonian manifestly:
\begin{eqnarray}
  H_{s}= \sum_{i,\sigma,\tau} \lambda\para{f_{i,A,\sigma}^\dag f_{i,A,\sigma}-F_{i,B,\sigma}^\dag F_{i,B,\sigma} } -t\sum_{i,\vec{\delta},\sigma,\tau}\Delta_{b}\para{i,j}\sigma f_{i,A,\sigma}^\dag F_{j,B,-\sigma} +h.c. ~.
\end{eqnarray}

In the absence of $\lambda$ energy band of spinons has two nodal points around
$\vec{K}=\frac{4\pi}{3\sqrt{3}a}\para{1,0}$ and
$\vec{K'}=\frac{4\pi}{3\sqrt{3}a}\para{\frac{1}{2},\frac{\sqrt{3}}{2}}$ (see Fig. 3).
If we expand the $\eta_{k}$ around these two
points we have:
\begin{eqnarray}
  &&\eta\para{K+\vec{q}}=\frac{3a}{2}\para{-q_x+iq_y}\\
  &&\eta\para{K'+\vec{q}}=\frac{3a}{2}\para{q_x+iq_y},
\end{eqnarray}
where $\vec{q}=\para{q_x,q_y}$. So we have 8
flavors of spinons depending on their physical spin degree of freedom,
sublattice index, and wether their momentum is around $K$ or $K'$. Therefore we
define the 8 component spinor as:
$\Psi^{\dag}\para{x}=\para{f_{A,K,\uparrow}^\dag\para{x},F_{B,K,\uparrow}^\dag\para{x},f_{A,K,\downarrow}^\dag\para{x}
,F_{B,K,\downarrow}^\dag\para{x},f_{A,K',\uparrow}^\dag\para{x}
,F_{B,K',\uparrow}^\dag\para{x},f_{A,K',\uparrow}^\dag\para{x},F_{B,K',\uparrow}^\dag\para{x}}$.
Using the linearized Hamiltonian around $K$ and $K'$ it is straightforward to
show that the continuum model can be written as:
\begin{eqnarray}
&&  H=\int d^2x ~\Psi^{\dag}\para{x}\left[\lambda \mu^3 \otimes \tau^0 \otimes \nu^0 -i \frac{3a}{2}\Delta_{b}\partial_{x}\mu^1 \otimes \tau^0 \otimes \nu^3 - i\frac{3a}{2}\Delta_{b}\partial_{y}\mu^2 \otimes \tau^0 \otimes \nu^0\right]\Psi\para{x},
\end{eqnarray}
where $\mu^{a}$ ($a=$0,1,2,3) are Pauli matrices acting on the sublattice
indices, $\tau^{b}$ are Pauli matrices acting on the physical spin, and
$\nu^{c}$ are Pauli matrices that act on the valley indices. It is known that
if filled band has a nontrivial total Chern number, then instanton effects can
be ignored, since the nontrivial Chern number lead to a Chern-Simons term for
the $U(1)$ gauge field.  So we need to calculate the Chern number of the filled
band.

We will calculate the Chern number through the Dirac nodes.  Each Dirac node
contribute $\pm 1/2$ Chern numbers, and each filled band have an even number of
Dirac nodes. So adding the contribution from all the Dirac nodes, we obtain an
integer Chern number for the filled band.

Let us consider the following two by two Dirac theory:
\begin{eqnarray}
  H=\int d^2x ~\psi^{\dag}\para{x}\left[m\sigma^3+\frac{3a}{2}\Delta_{b}\psi^{\dag}\para{x}\para{-i\sigma^1 \partial_{x}+i\sigma^2 \partial_{y}}\right] \psi\para{x},
\end{eqnarray}
where $\sigma^{i}$ are Pauli matrices. The mass of the above Hamiltonian is by definition $m$. It has been shown that each massive Dirac cone with mass $m$ has $C=\frac{m}{2\abs{m}}$ nontrivial Chern number \cite{Jackiw_1984_1,Jackiw_1998_1}. Now let us consider following Hamiltonian:
\begin{eqnarray}
  H=\int d^2x ~\psi^{\dag}\para{x}\left[m\sigma^3+\frac{3a}{2}\Delta_{b}\para{
-i\sigma^1 \partial_{x}-i\sigma^2 \partial_{y}}\right] \psi\para{x}.
\end{eqnarray}
If we us the following transformation: $\psi \rightarrow \psi'=\sigma^1 \psi$, then the above Hamiltonian can be rearranged as the following:
\begin{eqnarray}
  H=\int d^2x ~\psi'^{\dag}\para{x}\left[-m\sigma^3+\frac{3a}{2}\Delta_{b}\para{
-i\sigma^1 \partial_{x}+i\sigma^2 \partial_{y}}\right] \psi'\para{x}.
\end{eqnarray}
Therefore the mass of this Hamiltonian is $-m$, and therefore it has $C=-\frac{m}{2\abs{m}}$ nontrivial Chern number.


Using the above arguments it can be shown that the mass of the two Dirac cones
at $\vec{k}=\vec{K}$ is $\lambda<0$ and they contribute $C=-\frac{1}{2}$ Chern
number. The mass of the two other Dirac cones at $\vec{k}=\vec{K}'$ is
$-\lambda >0$ and they contribute $C=\frac{1}{2}$ Chern number. Therefore the
total Chern number of our theory is $2\times -1/2+2\times 1/2=0$. So the
coefficient of the Chern-Simon action is zero, and it does not constraint the
proliferation of instantons.

On the other hand since we have a massive Dirac
theory, instant-instanton correlation function at large distances DOES NOT
decay exponentially. Therefore nothing prevents instantons from proliferation.
They will proliferate and gap out the gauge particles.
So the $U(1)$ gauge theory is in the confined phase.
Now we should compute the quantum
numbers\cite{Goldstone_Wilczek_1981_1,Alicea_2008_1,Cenke_2008_1,Balents_Sachdev_2005_1,Balents_Sachdev_2005_2,Wingho_2009_1}
of instantons, in order to understand the symmetry properties of the $U(1)$
confined phase.  To do so let us first derive the instanton operators.

Since
instantons (instantons) in the presence of the Chern-Simon action with chern
number $C$, create $C$ fermions, therefore the instanton operator creates
$2\times -1/2=-1$ fermions at $\vec{k}=\vec{K}$ ({\em i.e.} annihilates 1 fermion at $\vec{K}$), and creates $2\times 1/2=1$
fermions at $\vec{k}=\vec{K}'$. Therefore, we obtain many different possibilities
for the instanton operators, which include:


\begin{eqnarray}
  &&\phi_{1}=f_{A,\uparrow,K'}^\dag f_{A,\uparrow,K}\\
  &&\phi_{2}=F_{B,\uparrow,K'}^\dag f_{A,\uparrow,K}=f_{B,\downarrow,K} f_{A,\uparrow,K}\\
  &&\phi_{3}=f_{A,\downarrow,K'}^\dag f_{A,\uparrow,K}\\
  &&\phi_{4}=F_{B,\downarrow,K'}^\dag f_{A,\uparrow,K}=f_{B,\uparrow,K} f_{A,\uparrow,K}\\
&&\cr
  &&\phi_{5}=f_{A,\uparrow,K'}^\dag F_{B,\uparrow,K}=f_{A,\uparrow,K'}^\dag f_{B,\downarrow,K'}^\dag\\
  &&\phi_{6}=F_{B,\uparrow,K'}^\dag F_{B,\uparrow,K}=f_{B,\downarrow,K} f_{B,\downarrow,K'}^\dag\\
  &&\phi_{7}=f_{A,\downarrow,K'}^\dag F_{B,\uparrow,K}=f_{A,\downarrow,K'}^\dag f_{B,\downarrow,K'}^\dag\\
  &&\phi_{8}=F_{B,\downarrow,K'}^\dag F_{B,\uparrow,K}=f_{B,\uparrow,K} f_{B,\downarrow,K'}^\dag\\
&&\cr
  &&\phi_{9}=f_{A,\uparrow,K'}^\dag f_{A,\downarrow,K}\\
  &&\phi_{10}=F_{B,\uparrow,K'}^\dag f_{A,\downarrow,K}=f_{B,\downarrow,K} f_{A,\downarrow,K}\\
  &&\phi_{11}=f_{A,\downarrow,K'}^\dag f_{A,\downarrow,K}\\
  &&\phi_{12}=F_{B,\downarrow,K'}^\dag f_{A,\downarrow,K}=f_{B,\uparrow,K} f_{A,\downarrow,K}\\
&&\cr
  &&\phi_{13}=f_{A,\uparrow,K'}^\dag F_{B,\downarrow,K}=f_{A,\uparrow,K'}^\dag f_{B,\uparrow,K'}^\dag\\
  &&\phi_{14}=F_{B,\uparrow,K'}^\dag F_{B,\downarrow,K}=f_{B,\downarrow,K} f_{B,\uparrow,K'}^\dag\\
  &&\phi_{15}=f_{A,\downarrow,K'}^\dag F_{B,\downarrow,K}=f_{A,\downarrow,K'}^\dag f_{B,\uparrow,K'}^\dag\\
  &&\phi_{16}=F_{B,\downarrow,K'}^\dag F_{B,\downarrow,K}=f_{B,\uparrow,K} f_{B,\uparrow,K'}^\dag.
\end{eqnarray}

It is obvious that all the above operators carry nonzero crystal momentum which
is equal to $\vec{K}'-\vec{K}$. Since the microscopic Hubbard Hamiltonian does
not break translation symmetry, therefore the single instanton operator cannot
appear in the path integral. On the other hand since
$3\para{\vec{K}'-\vec{K}}\equiv \para{0,0}$, triple-instanton is not forbidden
and will appear in the path integral.  So the the path integral contain a
triple-instanton gas, which will cause a confinement of the $U(1)$ theory.

\section{Quantum number of instantons}

\begin{table}
\label{tab1}
  \centering
  \begin{tabular}{|c| c| c| c| c| c|  }
  \hline
    $~~~$~~&~~~$T_1$~~~&~~~$T_2$~~~&~~$T$~~~&~~~$\sigma$~~~&~~~$C_6$\\
  \hline
$~~~f_{A,\alpha,K}\para{x}$~~&~~~$\exp\para{i\frac{4\pi}{3}}f_{B,\alpha,K}\para{T_1x}$~~~&~~$\exp\para{i\frac{2\pi}{3}}f_{B,\alpha,K}\para{T_2x}$~~~~&~~$\alpha f_{A,-\alpha,K'}\para{x}$~~~~&~~~~$ f_{B,\alpha,K}\para{\sigma x}$~~~&~~~$f_{B,\alpha,K'}\para{C_6x}$\\
\hline
$~~~f_{B,\alpha,K}\para{x}$~~&~~~$\exp\para{i\frac{4\pi}{3}}f_{A,\alpha,K}\para{T_1x}$~~~&~~$\exp\para{i\frac{2\pi}{3}}f_{A,\alpha,K}\para{T_2x}$~~~~&~~$\alpha f_{B,-\alpha,K'}\para{x}$~~~~&~~~~$ f_{A,\alpha,K}\para{\sigma x}$~~~&~~~$f_{A,\alpha,K'}\para{C_6x}$\\
\hline
$~~~f_{A,\alpha,K'}\para{x}$~~&~~~$\exp\para{i\frac{2\pi}{3}}f_{B,\alpha,K'}\para{T_1x}$~~~&~~$\exp\para{i\frac{4\pi}{3}}f_{B,\alpha,K'}\para{T_2x}$~~~~&~~$\alpha f_{A,-\alpha,K}\para{x}$~~~~&~~~~$ f_{B,\alpha,K'}\para{\sigma x}$~~~&~~~$f_{B,\alpha,K}\para{C_6x}$\\
\hline
$~~~f_{B,\alpha,K'}\para{x}$~~&~~~$\exp\para{i\frac{2\pi}{3}}f_{A,\alpha,K'}\para{T_1x}$~~~&~~$\exp\para{i\frac{4\pi}{3}}f_{A,\alpha,K'}\para{T_2x}$~~~~&~~$\alpha f_{B,-\alpha,K}\para{x}$~~~~&~~~~$ f_{A,\alpha,K'}\para{\sigma x}$~~~&~~~$f_{A,\alpha,K}\para{C_6x}$\\
  \hline
  \end{tabular}
  \caption{Symmetry transformation rules of spinon operators under translation, time reversal, parity and $\pi/3$ rotation.}
\end{table}

Now we like to compute other nontrivial quantum numbers of the above instanton operators. To do so, let us first comment on the transformation of the continuum wave-function. Using transformation rules in Table 1, we obtain following relations:

$ $

{\em Physical spin rotation around z axis by angle $\theta$}  :   $S_z \Psi~~~\rightarrow e^{i\theta \tau^3/2}\Psi \para{x}.$

$ $

{\em Physical spin rotation around y axis. by angle $\theta$} :   $S_y \Psi ~~~\rightarrow  e^{i\theta \mu^3 \otimes \tau^2/2}\Psi \para{x}.$

$ $

{\em Translation $T_1$ } :                                        $T_1 \Psi~~~\rightarrow e^{-i\frac{2\pi}{3}}\Psi \para{x'}.$

$ $

{\em Translation $T_2$ } :                                        $T_2 \Psi~~~\rightarrow e^{+i\frac{2\pi}{3}} \Psi \para{x'}.$

$ $

{\em Time reversal } :                                            $T \Psi~~~\rightarrow  i\tau^2 \otimes \nu^1\Psi \para{x'}.$

$ $

{\em Parity } :                                                   $\sigma \Psi~~~\rightarrow \mu^1 \otimes \tau^1 \otimes \nu^0\Psi^\dag\para{x'}.$

$ $

{\em $C_6$ } :                                                    $C_6 \Psi~~~\rightarrow \mu^1 \otimes \tau^1 \otimes \nu^1\Psi^\dag\para{x'}.$

$ $

\noindent
where $x'$ is the transformed $x$ under each symmetry transformations.

\subsection{Symmetry transformations on instanton operators}

Using symmetry transformation of continuum wavefunction, we can read the corresponding transformation of monopole operators. Under $\pi/3$ rotation ($C_6$) we have:
\begin{eqnarray}
&& C_6 \ \ \ : \ \ \  \phi_{i}\rightarrow -\phi_{17-i}^\dag.
\end{eqnarray}
Parity operator $\sigma$ where takes $y$ to $-y$ acts on monopole operators as following:
\begin{eqnarray}
&&  \phi_{1}\rightarrow -\phi_{16}, ~~~~~~~  \phi_{2}\rightarrow -\phi_{12}, ~~~~~~   \phi_{3}\rightarrow -\phi_{8}, ~~~~~~~  \phi_{4}\rightarrow -\phi_{4} \\
&&  \phi_{5}\rightarrow -\phi_{15}, ~~~~~~~ \phi_{6}\rightarrow -\phi_{11}, ~~~~~~  \phi_{7}\rightarrow -\phi_{7}, ~~~~~~~  \phi_{8}\rightarrow -\phi_{3}\\
&&  \phi_{9}\rightarrow -\phi_{14}, ~~~~~~~  \phi_{10}\rightarrow -\phi_{10}, ~~~~~  \phi_{11}\rightarrow -\phi_{6},~~~~~~  \phi_{12}\rightarrow -\phi_{2}\\
&&  \phi_{13}\rightarrow -\phi_{13}, ~~~~~~  \phi_{14}\rightarrow -\phi_{9},~~~~~~  \phi_{15}\rightarrow -\phi_{5},~~~~~~  \phi_{16}\rightarrow -\phi_{1}.
\end{eqnarray}

Under time reversal $T$ we have
\begin{eqnarray}
&&  \phi_{1}\rightarrow +\phi_{11}^\dag, ~~~~~~  \phi_{2}\rightarrow -\phi_{15}^\dag,~~~~~~~  \phi_{3}\rightarrow -\phi_{3}^\dag,~~~~~~~  \phi_{4}\rightarrow +\phi_{7}^\dag\\
&&  \phi_{5}\rightarrow -\phi_{12}^\dag, ~~~~~~\phi_{6}\rightarrow +\phi_{16}^\dag,~~~~~~~  \phi_{7}\rightarrow +\phi_{4}^\dag, ~~~~~~~  \phi_{8}\rightarrow -\phi_{8}^\dag\\
&&  \phi_{9}\rightarrow -\phi_{9}^\dag, ~~~~~~~  \phi_{10}\rightarrow +\phi_{13}^\dag,~~~~~~  \phi_{11}\rightarrow +\phi_{1}^\dag,~~~~~~  \phi_{12}\rightarrow -\phi_{5}^\dag\\
&&  \phi_{13}\rightarrow +\phi_{10}^\dag,~~~~~  \phi_{14}\rightarrow -\phi_{14}^\dag,~~~~~~  \phi_{15}\rightarrow -\phi_{2}^\dag,~~~~~~  \phi_{16}\rightarrow +\phi_{6}^\dag.
\end{eqnarray}
Monopole operators transform under translation along $\vec{R}_1$ as
\begin{eqnarray}
  && T_1 \ \ \ : \ \ \  \phi_{k}\rightarrow \mbox{exp}\para{i\frac{2\pi}{3}}\phi_{k}.
\end{eqnarray}
Similarly for translation along $\vec{R}_2$ we have
\begin{eqnarray}
  &&  T_2 \ \ \ : \ \ \ \phi_{k}\rightarrow \mbox{exp}\para{-i\frac{2\pi}{3}}\phi_{k}.
\end{eqnarray}

And finally rotation around z axis by $\theta$ angle:

\begin{eqnarray}
&&  \phi_{1}\rightarrow \phi_{1},~~~~~~~~~~~~~~~~~~~~  \phi_{2}\rightarrow \phi_{2},~~~~~~~~~~~~~~~~~~~~~  \phi_{3}\rightarrow \mbox{exp}\para{i\theta}\phi_{3},~~~~~~  \phi_{4}\rightarrow \mbox{exp}\para{i\theta}\phi_{4}\\
&&  \phi_{5}\rightarrow \phi_{5},~~~~~~~~~~~~~~~~~~~~  \phi_{6}\rightarrow \phi_{6},~~~~~~~~~~~~~~~~~~~~~  \phi_{7}\rightarrow \mbox{exp}\para{i\theta}\phi_{7},~~~~~~  \phi_{8}\rightarrow \mbox{exp}\para{i\theta}\phi_{8}\\
&&  \phi_{9}\rightarrow \mbox{exp}\para{-i\theta}\phi_{9},~~~~~~~  \phi_{10}\rightarrow \mbox{exp}\para{-i\theta}\phi_{10},~~~~~~  \phi_{11}\rightarrow \phi_{11},~~~~~~~~~~~~~~  \phi_{12}\rightarrow \phi_{12}\\
&&  \phi_{13}\rightarrow \mbox{exp}\para{-i\theta}\phi_{13},~~~~~ \phi_{14}\rightarrow \mbox{exp}\para{-i\theta}\phi_{14},~~~~~~  \phi_{15}\rightarrow \phi_{15},~~~~~~~~~~~~~~ \phi_{16}\rightarrow \phi_{16}.
\end{eqnarray}

The instanton operators should have the same symmetry as the microscopic Hamiltonian and therefore they carry trivial quantum numbers. Using the above transformations, it is easy to see The following term is invariant under all transformations:
\begin{eqnarray}
  L=g \int d^2 x \para{\phi^3\para{x}+\phi^{\dag 3}\para{x}},
\end{eqnarray}
where $\phi$ is defined as the following:
\begin{eqnarray}
 \phi&&=\phi_{2}-\phi_{12}+\phi_{5}-\phi_{15}\\
     &&=f_{B,\downarrow,K} f_{A,\uparrow,K}-f_{B,\uparrow,K} f_{A,\downarrow,K} +f_{A,\uparrow,K'}^\dag f_{B,\downarrow,K'}^\dag -f_{A,\downarrow,K'}^\dag f_{B,\uparrow,K'}^\dag.
\end{eqnarray}
This operator has the following symmetry properties:
\begin{eqnarray}
&&\sigma:~~~~~\phi~~\rightarrow~~\phi \\
&&T:~~~~~\phi~~\rightarrow~~\phi^\dag \\
&&C_6:~~~~\phi~~\rightarrow~~\phi^\dag \\
&&T_1:~~~~\phi~~\rightarrow~~\mbox{exp}\para{-i\frac{2\pi}{3}}\phi \\
&&T_2:~~~~\phi~~\rightarrow~~\mbox{exp}\para{+i\frac{2\pi}{3}}\phi.
\end{eqnarray}

It can be easily checked that $\phi$ operator is also invariant under rotation $x$ and $y$ axes.

\begin{figure}
\centering
\subfigure{
\includegraphics[height=110pt]{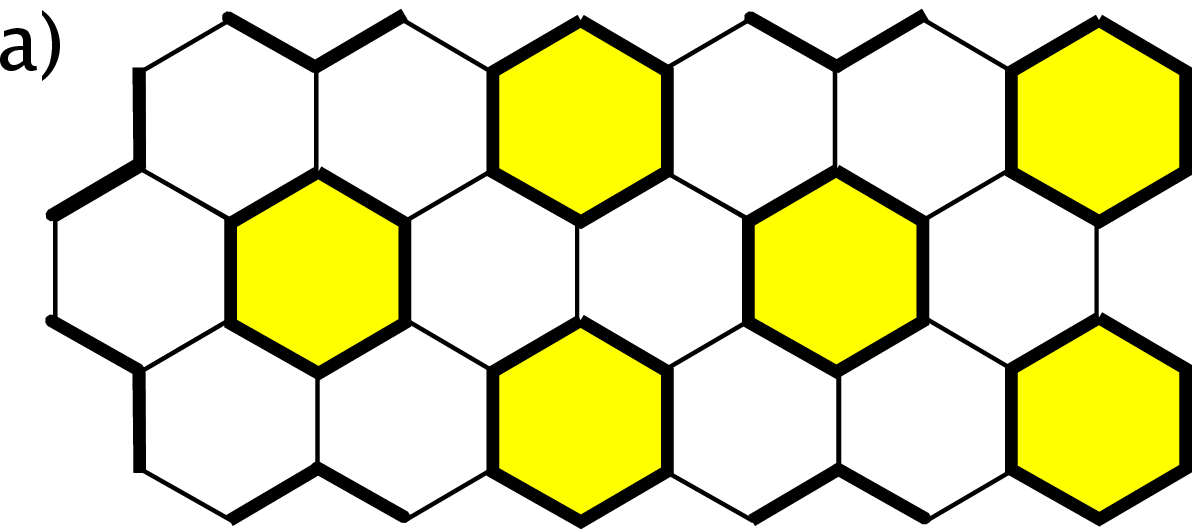}
\label{fig:subfig4a}
}
\subfigure{
\includegraphics[height=110pt]{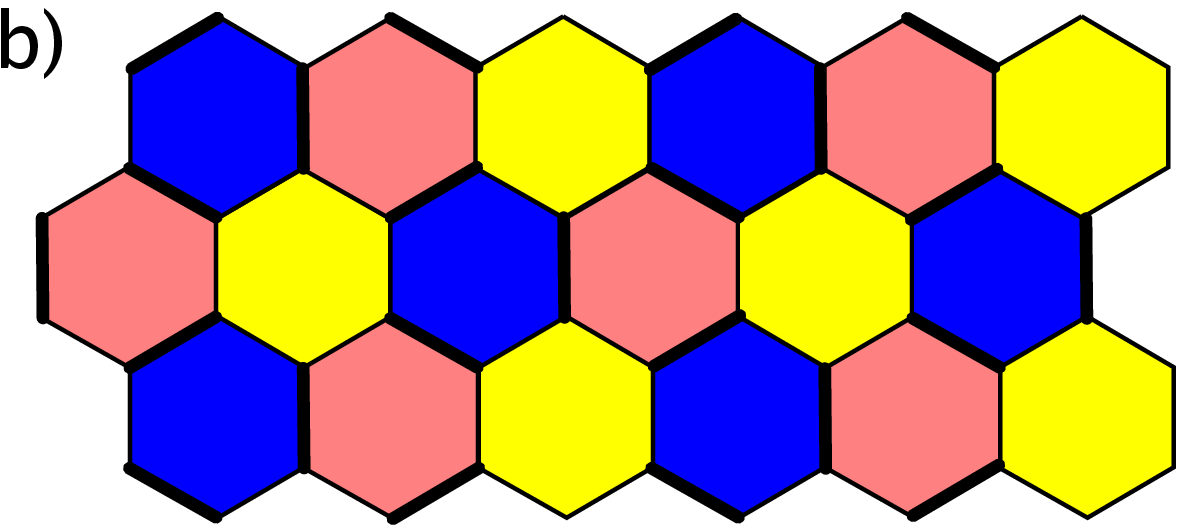}
\label{fig:subfig4b}
}
\label{fig:subfigureExample}
\caption[Optional caption for list of figures]{Two possible valence bond solid (VBS) states in honeycomb lattice that break the translation symmetry. Bold line indicates the stronger bonds and narrow line the weaker links. The bond operator in our case corresponds to the exchange energy {\em i.e.} $\Delta_{b}\para{i,j}$. $\vec{R}'_{1}=3\vec{R}_1$ and $\vec{R}'_{2}=\vec{R}_1 +\vec{R}_2 $ are new basis vectors of the lattice. The area of the unit cell is three times bigger than the translation symmetric case and contains six atoms in it. \subref{fig:subfig4a}, Honeycomb lattice with broken translation symmetry, while $C_6$ and time reversal symmetries are unbroken. This phase corresponds to $\theta=0$. $\theta$ can correspond to the flux of the hexagon. \subref{fig:subfig4b}, Honeycomb lattice with broken translation, $C_6$ rotation and time reversal symmetry (center of rotation is yellow hexagons). This state corresponds to $\theta=\frac{2\pi}{3}$. The phase with $\theta=-\frac{2\pi}{3}$ is related to this by $C_6$ or $T$. From this figure it is clear that in this case $C_6$ breaks down to $C_3$.}\label{fig4}
\end{figure}

\section{Discussion and conclusion}

We have found a triple-instanton operator that has all symmetries of the
microscopic Hamiltonian. Therefore this term is relevant and has a non-zero
fugacity. Because of instanton proliferation, the U(1) gauge fluctuations are
now gapped out. On the other hand, since single instanton operator carries
nonzero crystal momentum, the translation symmetry breaks spontaneously. To
have a better insight of the situation, we can use the duality between $U(1)$
gauge theory and the nonlinear sigma model. If we identify $\phi$ operator with
$\mbox{exp}\para{i\theta}$, then $g \para{\phi^3\para{x}+\phi^{\dag
3}\para{x}}=2g \cos\para{3\theta}$. The $U(1)$ gauge theory
with triple-instantons can be described by the following dual theory:
\begin{eqnarray}
  L=\frac{m}{2}\dot{\theta}^2-\frac{\rho}{2} \para{\nabla{\theta}}^2+2g\cos\para{3\theta}.
\end{eqnarray}
Therefore $T$ and $C_6$ transformation are equivalent to $\theta \rightarrow
-\theta$, $\sigma$ is trivial and $T_1$ and $T_2$ are equivalent to $\theta
\rightarrow \theta -\frac{2\pi}{3} $ and $\theta \rightarrow \theta
+\frac{2\pi}{3} $ respectively. This model has three inequivalent ground states
determined for $\theta \in \left\{0,\frac{2\pi}{3},\frac{4\pi}{3}\right\}$ (see Fig. 4).
Therefore the ground-state degeneracy of our model is also three. This happens
because we lose the translation symmetry along $\vec{a}_1$ and $\vec{a}_2$
direction. The reason is that $\phi$ operator carries nonzero crystal momentum.
But it is easy to show that $\phi$ is invariant under $T_1^3$ and $T_1 T_2$
transformation. So the basis vectors of the new lattice are
$\vec{R}'_1=\para{3,0}=3\vec{R}_1$ and $\vec{R}'_2=\para{1,1}=\vec{R}_1+\vec{R}_2$. The area of the
unit cell is three times bigger and it contains six atoms in it.

In summary, if we treat the U(1) gauge field as a semi-classical field ({\em
i.e.}Gaussian approximation), by analogy to the nonlinear sigma model,
instantons proliferate, gauge field gaps out and lattice symmetry spontaneously
breaks.  What we finally obtain is a band insulator instead of a spin liquid
phase. We want to mention that if gauge fluctuations are strong, other
possibilities may happen. Among them, the $Z_2$ spin liquid is of more
interest. This phase can be obtained if strong gauge fluctuations generate
hopping term to the nearest site or a nonzero pairing amplitude to the second
neighboring site. The presence of any of these two terms breaks U(1) down to
$Z_2$ gauge theory which is stable.




\section*{Acknowledgement} We thank P.A. Lee, T. Senthil, B. Swingle and F.
Wang  for their useful comments and helpful discussions. This research is
supported by  NSF Grant No. DMR-1005541 and NSFC 11074140.

\end{document}